\documentclass{PoS}

\PoS{PoS(HEP2005)108}

\title{On a new unitarization scheme inspired by Dalitz and Tuan applied to meson-meson and meson-baryon scattering}

\ShortTitle{On a new unitarization scheme inspired by Dalitz and Tuan $\ldots$}

\author{\speaker{Frieder Kleefeld}\thanks{This work has been supported by the
{\em Funda\c{c}\~{a}o para a Ci\^{e}ncia e a Tecnologia} \/(FCT) of the {\em Minist\'{e}rio da Ci\^{e}ncia, Tecnologia e Ensino Superior} \/of Portugal, under Grants no.\ PRAXIS XXI/BPD/20186/99, SFRH/BDP/9480/2002, POCTI/\-FNU/\-49555/\-2002, and POCTI/FP/FNU/50328/2003.}\\
        Centro de F\'{\i}sica das Interac\c{c}\~{o}es Fundamentais (CFIF),\\
        Instituto Superior T\'{e}cnico,\\
        Av.\ Rovisco Pais, 1049-001 Lisboa, Portugal\\
        E-mail: \email{kleefeld@cfif.ist.utl.pt}}

\abstract{A new crossing symmetric unitarization scheme conveniently applied to meson-meson and meson-baryon scattering amplitudes is shortly proposed which can be not only used by theoreticians to unitarize arbitrary theoretical reaction amplitudes resulting from phenomenological Lagrangeans for mesons and baryons, yet also by experimentalists to generate realistic unitary fitting formulae for meson-meson and meson-baryon scattering observables sharing on one hand all the features of the underlying theoretical amplitudes, on the other hand allowing direct comparison to these amplitudes. The new unitarization scheme has been inspired by the Dalitz and Tuan (DT) representation \cite{Dalitz:1960}, the basic ansatz of which is that {\em ``... the phases caused by different sources add ...''} (using the words of B.S.\ Zou, D.V.\ Bugg, Phys.\ Rev.\ D 50 (1994) 591 \cite{Zou:1994ea}).}

\FullConference{International Europhysics Conference on High Energy Physics\\
		July 21st - 27th 2005\\
		Lisboa, Portugal}

\begin{document}

\section{Unitarization and a new unitarization scheme inspired by Dalitz and Tuan}
\begin{figure}[t]
\begin{minipage}[t]{0.32\linewidth}
\begin{center}
\includegraphics[width=1.0\textwidth]{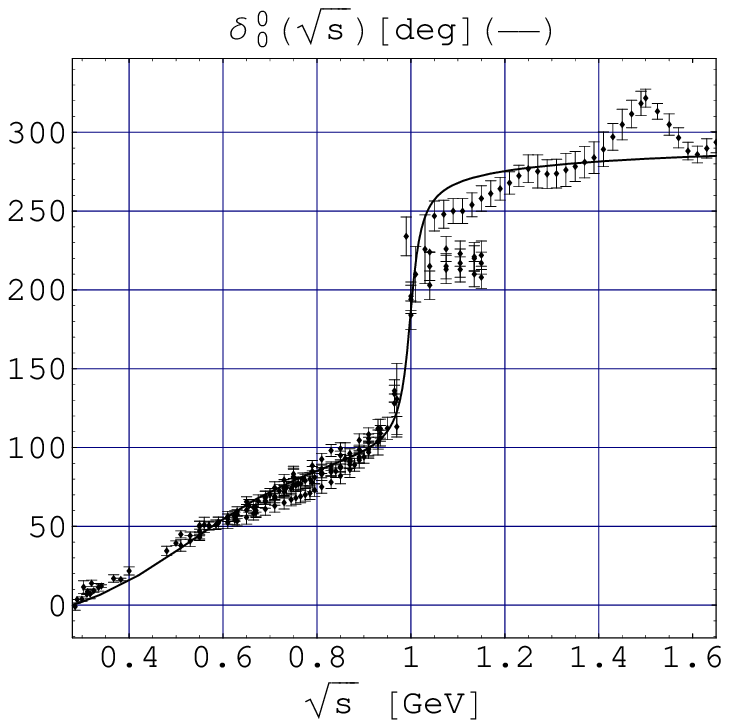} \\
\end{center}
\caption{Fit of $\delta^0_0(\sqrt{s})$: contribution of ``background'' ($-67.5^\circ$) and $\sigma(600)$ and $f_0(980)$.
}
 \label{fig1}
\end{minipage}%
\hspace{0.015\textwidth}%
\begin{minipage}[t]{0.32\linewidth}
\begin{center}
\includegraphics[width=1.0\textwidth]{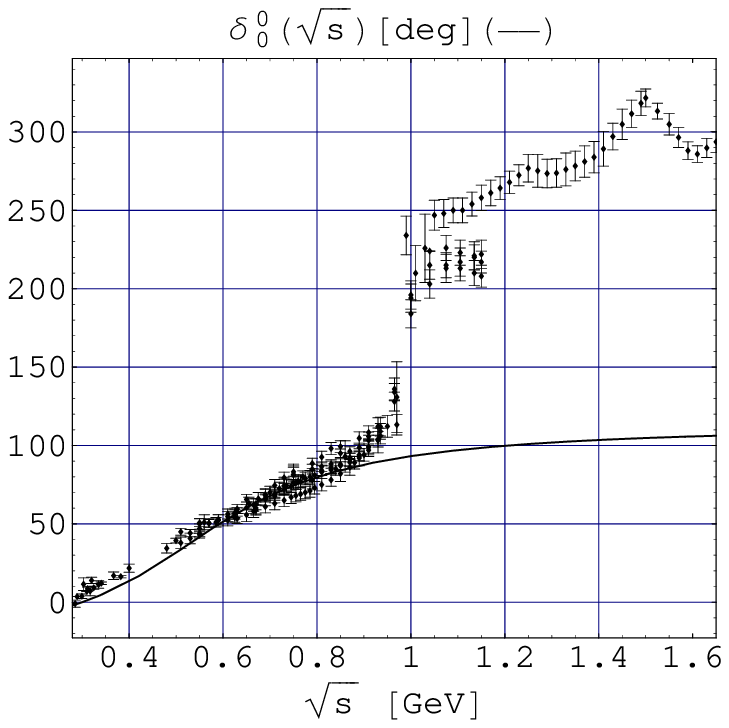} \\
\end{center}
\caption{Fit of $\delta^0_0(\sqrt{s})$: contribution of ``background'' ($-67.5^\circ$) and $\sigma(600)$.
}\label{fig2}
\end{minipage}
\hspace{0.015\textwidth}
\begin{minipage}[t]{0.32\linewidth}
\begin{center}
\includegraphics[width=1.0\textwidth]{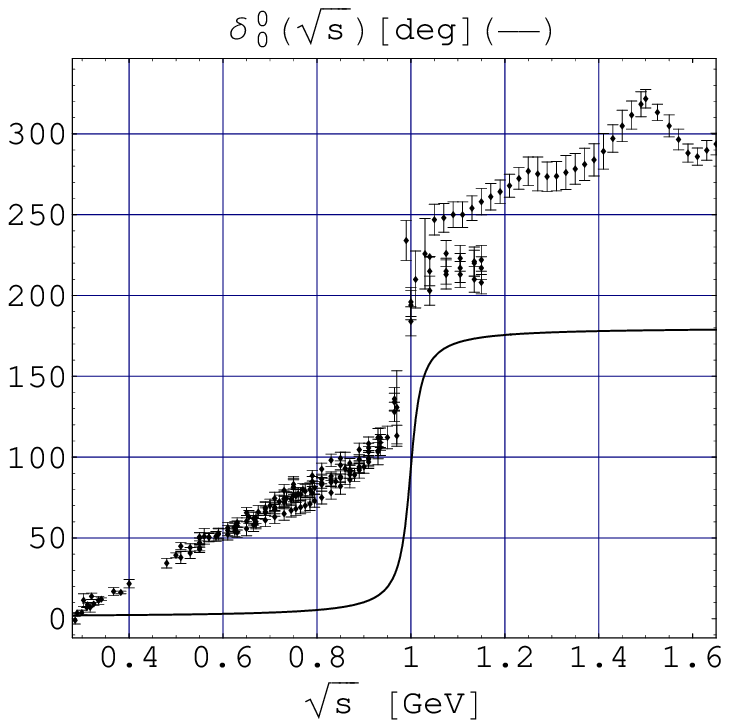} \\
\end{center}
\caption{Fit of $\delta^0_0(\sqrt{s})$: contribution of $f_0(980)$. 
}
 \label{fig3}
\end{minipage}%
\end{figure}

A quantitative description of particle scattering/production processes involving strong interactions requires the complete non-perturbative scattering/production amplitude even close to threshold. We shall call the procedure of estimating the non-perturbative part of a scattering/production amplitude on the basis of a known perturbative or ``tree-level'' amplitude {\em unitarization} \cite{Kleefeld:2005qs}. A comprehensive list of existing unitarization methods has been provided and discussed in Ref.\ \cite{Truong:1991gv}: the Pad\'{e} method \cite{Basdevant:1972,Truong:1988zp}, the Inverse Amplitude Method \cite{Truong:1991gv,Tryon:1971pt,Dobado:1996ps,Truong:1988zp}, the N/D method, and K-matrix unitarization method \cite{Zou:1994ea,Truong:1991gv,Chung:1995dx}. Below we propose a to our best knowledge new \emph{crossing-symmetric} unitarization method that avoids drawbacks\footnote{Serious drawbacks of commonly used unitarization methods are lack of crossing symmetry (e.g.\ \cite{Cavalcante:2003iq}), truncation dependences \cite{Dobado:1996ps}, problems with inclusion of chiral zeros \cite{Dobado:1996ps}, and difficulties to relate analytic expressions resulting from such methods directly to results obtained by quantum-field theoretic calculations of scattering amplitudes on the basis of Lagrangeans. 
Roy-equations \cite{Roy:1971tc} admit fortunately crossing symmetry by imposing external conditions, unfortunately they make strong assumptions about analyticity and have to be truncated in order to allow a numerical solution.} of above mentioned unitarization schemes and is inspired by the observation of Dalitz and Tuan  (DT) \cite{Dalitz:1960} that --- using the words of B.S.\ Zou and D.V.\ Bugg  \cite{Zou:1994ea} --- {\em ``... the phases caused by different sources add ..."}. 
\footnote{Starting point for our consideration has been in analogy to DT the observation that the S-matrix for scalar isoscalar $\pi\pi\rightarrow \pi\pi$-scattering at energies below the $K\bar{K}$-threshold and slightly above the $\pi\pi$-threshold is fitted on the basis of experimental pole positions $M_{\sigma(600)}=(0.525 - i\,0.265)$~GeV and $M_{f_0(980)}=(0.999 - i\,0.017)$~GeV to an astonishing good approximation by a product of partial (Breit-Wigner) S-matrices $S_{\sigma(600)}(s)=(s - M^{\ast \, 2}_{\sigma(600)})/(s - M^2_{\sigma(600)})$ and $S_{f_0(980)}(s)=(s - M^{\ast \, 2}_{f_0(980)})/(s - M^2_{f_0(980)})$ and a background phase $S^\ast_{\sigma(600)}(s_{thr})\,S^\ast_{f_0(980)}(s_{thr})=\exp \left(2 i\,(-\,67.5^\circ)\right)$ being suitably chosen to make the cross section vanish exactly at the $\pi\pi$-threshold $s_{thr} = 4\,m^2_{\pi^\pm}$ (with $m_{\pi^\pm} = 0.13957018$~GeV). The fit $S(s) = \exp \left(2 i\,(-\,67.5^\circ)\right) \, S_{\sigma(600)}(s)\, S_{f_0(980)}(s)$ --- lacking by construction desirable square-root behaviour at $s=s_{thr}$ --- is compared to experimental phaseshift data in Figs.\ \ref{fig1}, \ref{fig2} and \ref{fig3}. $S_{\sigma(600)}(s)$ and $S_{f_0(980)}(s)$ represent manifestly s-channel $\pi\pi$-scattering, while the background phase $-\,67.5^\circ$ carries the reminder of $t$- and $u$-channel scattering processes.} 
Hence, we assume the S-matrix $S$ to be factorizable as a product of several partial S-matrices $S_1$, $\ldots$, $S_n$ ($n\in I\!\!N$), i.e.\ $S = S_1 S_2  \ldots  S_n$ ($n\in I\!\!N$). For simplicity we could assume unitarity of the partial S-matrices yielding the relation $S_j = 1 + 2\, i \, \bar{T}_j$ ($j=1,\ldots,n$) between partial S-matrices and corresponding partial T-matrices $\bar{T}_1$,~$\ldots$,~$\bar{T}_n$.
Then the T-matrix is expanded into partial T-matrices $\bar{T}_1$, $\ldots$, $\bar{T}_n$ as follows:
\begin{equation} T \; = \; \frac{(S - 1)}{2\, i} \; = \; \frac{1}{2\, i} \left( \prod\limits^{n}_{j=1} S_j \; - \;  1\right) \; = \; \underbrace{\sum\limits^{n}_{j=1} \; \bar{T}_j}_{\mbox{``tree-level''}} + \underbrace{\frac{1}{2\, i} \left( \prod\limits^{n}_{j=1} \; ( 1 + 2\, i \; \bar{T}_j ) \; - \;  1\right) - \sum\limits^{n}_{j=1} \; \bar{T}_j}_{\mbox{``unitarization correction''}} \; .
\end{equation}
An instructive case occurs when all $n$ partial T-matrices are equal to one partial T-matrix $\bar{T}\equiv \bar{T}_1 = \bar{T}_2 = \ldots = \; \bar{T}_n$ corresponding to an unitary partial S-matrix $1+ 2\,i \, \bar{T}$. Then we have $S =  (1+ 2\,i \, \bar{T})^n$ and $T = (S - 1)/(2\, i) = ((1+ 2\,i \, \bar{T})^n - 1)/(2\, i)  = n\, \bar{T} +$ ``unitarization correction''.
Hence the $n$-th power of a partial S-matrix $1+ 2\,i \; \bar{T}$ yields a ``tree-level'' T-matrix $n\, \bar{T}$ being $n$ {\em times} the corresponding partial T-matrix $\bar{T}$. \footnote{It is straight forward to analytically continue this result to {\em arbitrary rational} values of $n$. For  $n\in I\!\!R\backslash I\!\!N_0$ we obtain:
\begin{equation}
 T \; = \;  \frac{(1+ 2\,i \; \bar{T})^n - 1}{2\, i} \; = \; n\; \bar{T} \; (\mbox{``tree-level''}) + \frac{1}{2\,i} \sum\limits^\infty_{j=2} \; \frac{n\, (n-1) \, \cdots \, (n-j+1)}{j!} \; (2\,i\; \bar{T})^{j}\;(\mbox{``unitarization corr.''}).
\end{equation}
A simple non-trivial example for a non-natural value of $n$ is $n=-1$, displaying a strong similarity to K-matrix unitarization, as $S=(1+ 2\,i \; \bar{T})^{-1}$ and therefore $T=(S-1)/(2i)=- \bar{T}/(1+ 2\,i \; \bar{T})$.}
Let's proceed to the case, in which we don't know, whether a partial T-matrix $\bar{T}_{j}$ ($j\in I\!\!N$) corresponds to an unitary partial S-matrix $S_{j}$ ($j\in I\!\!N$) or not. Inspired by DT we would at least expect that the phase of $\bar{T}_{j}$ determines the phase of $S_{j}$ according to $S_{j} = ( \bar{T}_{j}/\bar{T}^\ast_{j} )^{\alpha_j} \equiv ( 1 + 2 i \, \mbox{Im} [\bar{T}_{j}] \, \bar{T}^{\ast \,-1}_{j})^{\alpha_j}$, while $\alpha_j$ is determined such that the ``tree-level'' contribution to the T-matrix is given by $\bar{T}_{j}$ itself. To find $\alpha_j$ we expand $(S_{j}-1)/(2\,i)$ in terms of $\bar{T}_{j}$:
\begin{equation} \frac{S_{j}-1}{2\,i}
 = \underbrace{\alpha_j\; \frac{\mbox{Im} [\bar{T}_{j}]}{\bar{T}^\ast_{j}}}_{\mbox{``tree-level''}} + \underbrace{\frac{1}{2\,i} \sum\limits^\infty_{\ell =2} \; \frac{1}{\ell!} \; \left(2\,i\; \alpha_j\; \frac{\mbox{Im} [\bar{T}_{j}]}{\bar{T}^\ast_{j}}\right)^{\ell}\prod^{\ell}_{k=1}\left(1-\frac{k -1}{\alpha_j}\right)}_{\mbox{``unitarization correction''}} . 
\end{equation} 
Simple inspection yields $\alpha_j\, \mbox{Im} [\bar{T}_{j}]/\bar{T}^\ast_{j}  = \bar{T}_{j} \Rightarrow \alpha_j = |\bar{T}_{j}|^2/\mbox{Im} [\bar{T}_{j}]$. \footnote{The choice implies of course $\frac{S_{j}-1}{2\,i} = \bar{T}_{j}\;(\mbox{``tree-level''})
 + \frac{1}{2\,i} \sum\limits^\infty_{\ell =2} \; \frac{\left(2\,i\; \bar{T}_{j}\right)^{\ell}}{\ell !} \; \prod^{\ell}_{k=1}\left(1-(k -1)\frac{\displaystyle\mbox{Im} [\bar{T}_{j}]}{\displaystyle|\bar{T}_{j}|^2}\right)\;(\mbox{``unit.\ corr.''})$.}
Hence we conclude:
{\em If the ``tree-level'' T-matrix is given by a sum $\bar{T}_1 + \ldots +\bar{T}_n$ ($n\in I\!\!N$) of arbitrary partial T-matrices $\bar{T}_1$, $\ldots$, $\bar{T}_n$, then a DT-unitarized S-matrix $S$ with a ``tree-level'' term $\bar{T}_1 + \ldots +\bar{T}_n$ can be denoted as:}
\begin{eqnarray} S & = & S_1\; S_2 \; \ldots \; S_n \; = \; \left( \frac{\bar{T}_1}{\bar{T}^\ast_1} \right)^{\frac{\displaystyle|\bar{T}_1|^2}{\displaystyle\mbox{Im} [\bar{T}_1]}} \; \left( \frac{\bar{T}_2}{\bar{T}^\ast_2} \right)^{\frac{\displaystyle|\bar{T}_2|^2}{\displaystyle\mbox{Im} [\bar{T}_2]}} \; \ldots \; \left( \frac{\bar{T}_n}{\bar{T}^\ast_n} \right)^{\frac{\displaystyle|\bar{T}_n|^2}{\displaystyle\mbox{Im} [\bar{T}_n]}} \; = \nonumber \\
 & = & \exp \Bigg[\, i \Bigg( \frac{|\bar{T}_1|^2}{\mbox{Im} [\bar{T}_1]} \; \arg \left( \frac{\bar{T}_1}{\bar{T}^\ast_1} \right)+\frac{|\bar{T}_2|^2}{\mbox{Im} [\bar{T}_2]} \; \arg \left( \frac{\bar{T}_2}{\bar{T}^\ast_2} \right)+ \ldots +\frac{|\bar{T}_n|^2}{\mbox{Im} [\bar{T}_n]} \; \arg \left( \frac{\bar{T}_n}{\bar{T}^\ast_n} \right)  \Bigg) \Bigg]  . \;\; \quad  
\end{eqnarray}
\section{Instructive examples}
{\bf DT-unitarized s-channel Breit-Wigner resonance.} Consider a partial T-matrix for one s-channel Breit-Wigner resonance with constant complex mass $M$ dressed by a {\em real} coupling constant $g\in I\!\!R$. I.e.\ we make the ansatz $\bar{T}_1 = g\, \mbox{Im} [M^2]/(s-M^2)$ yielding $\mbox{Im}[\bar{T}_1] =  g\, \mbox{Im}^2 [M^2]/|s-M^2|^2$ and $|\bar{T}_1|^2  =  g^2\, \mbox{Im}^2 [M^2]/|s-M^2|^2$. Then the unitarized S-matrix inspired by DT will be given by $S = \left( \frac{s-M^{\ast 2}}{s-M^2} \right)^{g} = \exp \left[\, i  \; g \; \arg \left( \frac{s-M^{\ast 2}}{s-M^2} \right) \right]$. The resulting DT-unitarized T-matrix is given by:
\begin{equation} T\; = \; \frac{S-1}{2\,i} \; = \; \underbrace{g\; \frac{\mbox{Im} [M^2]}{(s-M^2)}}_{\mbox{``tree-level''}}  + \underbrace{\frac{1}{2\,i} \sum\limits^\infty_{\ell =2} \; \frac{g\, \left(g-1\right)\, \left(g-2 \right) \, \cdots \, \, \left(g-\ell +1\right)}{\ell !} \; \left(2\,i\; \frac{\mbox{Im} [M^2\,]}{(s-M^2)}\right)^{\ell}}_{\mbox{``unitarization correction''}} \; .
\end{equation}
In agreement with our previous discussion the unitarized T-matrix $T$ reduces to the ``tree-level'' T-matrix $T_1$ for $g\in I\!\!N_0$. $T$ seems to possess the same poles and zeros as $T_1$.

{\bf DT-unitarized one-channel two-resonance L$\sigma$M approach to $\pi\pi$-scattering}. The ``tree-level'' $U(3)\times U(3)$ Linear Sigma Model (L$\sigma$M) scattering amplitude for $\pi\pi\rightarrow \pi\pi$ scattering with $\sigma(600)$ and $f_0(980)$ intermediate states is given by ($f_\pi \simeq 92.42$ MeV): \footnote{$\phi_s\in I\!\!R$ is a nonstrange-strange scalar mixing angle defined by $\left| \sigma\right>=\cos \phi_s \left| n\bar{n}\right> - \sin \phi_s \left| s\bar{s}\right>$ and $\left| f_0\right>=\sin \phi_s \left| n\bar{n}\right> + \cos \phi_s \left| s\bar{s}\right>$, while $\rho_{\pi\pi}(s)= |\vec{p}^{\,\,\pi\pi}_{cm}|/(8\pi \, \sqrt{s}) =  \sqrt{(1 - 4\, m^2_\pi \, s^{-1})}/(16 \pi) \simeq \theta(s - 4\, m^2_\pi)/(16 \pi)$ is the $\pi\pi$-phasespace.}
\begin{eqnarray} \lefteqn{\bar{T}_{\pi^0\pi^0\leftarrow\pi^+\pi^-}(s,t,u) \; = \; \sqrt{\rho_{\pi\pi}(s)} \;  \frac{s - m^2_\pi}{f^2_\pi} \; \Bigg( 1- (s - m^2_\pi) \Bigg( \frac{\cos^2\phi_s}{s-M^2_\sigma} + \frac{\sin^2\phi_s}{s-M^2_{f_0}} \Bigg)\Bigg) \sqrt{\rho_{\pi\pi}(s)}\;=} \nonumber \\
 & = & \sqrt{\rho_{\pi\pi}(s)}  \Bigg( \frac{s - m^2_\pi}{f^2_\pi}   -   \frac{(s - m^2_\pi)^2}{f^2_\pi\; \mbox{Im}[M^2_\sigma]} \, \cos^2\phi_s \, \frac{\mbox{Im}[M^2_\sigma]}{s-M^2_\sigma} -  \frac{(s - m^2_\pi)^2}{f^2_\pi \; \mbox{Im}[M^2_{f_0}]} \, \sin^2\phi_s \,\frac{\mbox{Im}[M^2_{f_0}]}{s-M^2_{f_0}} \Bigg)  \sqrt{\rho_{\pi\pi}(s)}\, . \nonumber  
\end{eqnarray}
The DT-unitarized crossing symmetric S-matrix is now obviously ($P_{I=0,1,2}$ are isospin projectors)
\begin{eqnarray} \lefteqn{S_{\pi\pi\leftarrow \pi\pi}(s,t,u) \; = \; \exp \Big[ \, i \; \sqrt{\rho_{\pi\pi}(s)} \; \times} \nonumber \\
 & \times & \Big( 3  P_{I=0}\, (\Delta(s) +\Delta(t) +\Delta(u)) + P_{I=1}\, (\Delta(t) -\Delta(u)) + P_{I=2} \, (\Delta(t) +\Delta(u))\Big) \sqrt{\rho_{\pi\pi}(s)} \, \Big]
\end{eqnarray}
with $\Delta(s) \equiv  - \,  \frac{(s - m^2_\pi)^2}{f^2_\pi\; \mbox{Im}[M^2_\sigma]} \; \cos^2\phi_s \; \arg \left( \frac{s-M^{\ast\, 2}_\sigma}{s-M^2_\sigma} \right) - \, \frac{(s - m^2_\pi)^2}{f^2_\pi \; \mbox{Im}[M^2_{f_0}]} \; \sin^2\phi_s \;\arg \left( \frac{s-M^{\ast\, 2}_{f_0}}{s-M^2_{f_0}} \right)$.

%  Fig. 1,   Fig. 1

\end{document}